\documentclass[a4paper,12pt]{article}
\DeclareMathSizes{12}{12.5}{10}{10}
\usepackage[left=2.3cm,bottom=3cm,right=2.3cm,top=3cm]{geometry}
\usepackage{overpic,youngtab}
\usepackage{subfigure}
\usepackage[latin,english]{babel}
\usepackage{amsmath}
\usepackage{verbatim}
\usepackage{amssymb}
\usepackage{latexsym}
\usepackage{epsfig}
\usepackage{epstopdf}
\usepackage{scalerel}
\usepackage{graphics,psfrag,rotating}
\usepackage{graphicx}
\usepackage{dcolumn}
\usepackage{lscape}
\usepackage{float}
\usepackage{pdflscape}
\usepackage{array}
\usepackage{booktabs}
\usepackage{amscd}
\usepackage{fancybox}
\usepackage{color}
\usepackage[colorlinks=true,citecolor=blue,,linktocpage=true,linkcolor=blue,urlcolor=black]{hyperref}
\usepackage{mathabx}

\definecolor{ogreen}{rgb}{0,0.7,0}

\DeclareGraphicsExtensions{.pdf,.png,.jpg,.gif,.jpeg}
\def\be{\begin{equation}}
\def\ee{\end{equation}}
\def\bea#1\eea{\begin{align}#1\end{align}}
\def\pd{\partial}
\def\a{\alpha}
\def\b{\beta}
\def\g{\gamma}
\def\d{\delta}
\def\e{\epsilon}

\def\m{\mu}
\def\n{\nu}

\def\t{\tau}

\def\l{\lambda}

\def\r{\rho}

\def\s{\sigma}
\def\e{\epsilon}
\def\t{\tau}
\def\bi{\begin{itemize}}
	\def\ei{\end{itemize}}

\usepackage{authblk}
\usepackage{setspace}

\begin{document}
		\vspace*{-1cm}
	{\flushleft
		{{FTUAM-18-26}}
		\hfill{{IFT-UAM/CSIC-18-111}}}
	\vskip 1.5cm
	\begin{center}
		{\LARGE\bf A note on quasi-local energy}\\[3mm]
		\vskip .3cm
		
	\end{center}
	\vskip 0.5  cm
	\begin{center}
		{\Large Enrique Alvarez}$^{~a}$, {\Large Jesus Anero}$^{~a}$, {\Large Guillermo Milans del Bosch}$^{~a}$ {{\large and} \Large Raquel Santos-Garcia}$^{~a}$
		\\
		\vskip .7cm
		{
			$^{a}$Departamento de F\'isica Te\'orica and Instituto de F\'{\i}sica Te\'orica (IFT-UAM/CSIC),\\
			Universidad Aut\'onoma de Madrid, Cantoblanco, 28049, Madrid, Spain\\
			\vskip .1cm

			\vskip .5cm
			\begin{minipage}[l]{.9\textwidth}
				\begin{center} 
					\textit{E-mail:} 
					\tt{enrique.alvarez@uam.es},
					\tt{jesusanero@gmail.com},
				\tt{guillermo.milans@csic.es},
					\tt{raquel.santosg@uam.es}.
				\end{center}
			\end{minipage}
		}
	\end{center}
	\thispagestyle{empty}
	
	\begin{abstract}
		\noindent
		A definition of quasi-local energy in a gravitational field based upon its embedding into flat space is discussed. The outcome is not satisfactory from many points of view.
					\end{abstract}
		\newpage
	\tableofcontents

\flushbottom
\thispagestyle{empty}
	\newpage
	\setcounter{page}{1}
\section{Introduction}
One of the main problems in the analysis of the gravitational interaction is the non-existence of a natural definition of energy of the gravitational field itself. This can be argued from the equivalence principle: the energy-momentum cannot be a four-vector, because it can always be made to vanish locally in a free falling frame. This is at the root of many peculiarities such as ambiguities in the definition of the ground state of the gravitational field, and also the generalization of Schr\"odinger's equation\footnote{It is well known that the Wheeler-deWitt equation does not solve this particular problem, among other things, because there is no natural definition of time associated to it.} to the gravitational field (confer \cite{Alvarez} and references therein).  Some progress can be made in asymptotically flat spaces, but this is by no means satisfactory enough.
\par

A natural way out to this problem is to introduce a ``quasi-local" notion of energy (see \cite{Szabados} for a recent review), namely {\em quasi-local energy} (QLE). This provides a tool to compare spacetimes with different assymptotics \cite{AAMS}. A lot of effort has been made in the subject, leading to various definitions of QLE \cite{Hawking, Penrose, Bartnik,Brown:1992br, Hayward,Liu,Wang}. The QLE of the gravitational field contained in some region of the spacteime is defined as the integral of the extrinsic curvature over the hypersurface defining that region. Schematically,
\be
Q(\Sigma)\equiv \int_\Sigma K-E_0,
\ee
The main differences in the various definitions of the QLE come from the choice of the hypersurface and from the substraction of the zero-point energy $E_0$, which is computed by an isometric embedding of the hypersurface in $\mathbb{R}^3$ in the case of \cite{Brown:1992br}, or in $M_4$ in \cite{Wang}.
\par
The aim of the present article is to put forward a bold proposal. Namely to present a new idea to {\em define} the energy associated to an arbitrary gravitational field as embodied in a 4-dimensional metric. To that end, we use the Gauss-Codazzi equations to construct an action in a higher dimensional flat spacetime, where the embedding functions are the dynamical fields. The variation of this action yields a conserved energy momentum tensor, which from the 4-dimensional point of view, could potentially allow to define in a more intrinsic way he QLE of the gravitational field. Our initial purpose was to dispose of the usual  substraction of the  zero point energy as is usually done in the standard formulation of QLE in \cite{Brown:1992br, Wang}, as we are trying to define unambiguous quantities. We shall comment later to what extent we have suceeded in our aim.
\par
There are many issues to consider. We shall refer all the time to the embedding as defining a codimension $n-d$ hypersurface $\Sigma_d$ in a flat Minkowski n-dimensional  space, $M\equiv \mathbb{R}^{1,n-1}$, or else $\mathbb{R}^{2,n-2}$, with metric which will be denoted in either case by $\eta_{\m\n}$ in cartesian coordinates.
As for the existence of the necessary embedding, it is known \cite{Friedman} that any pseudo-riemannian n-dimensional manifold, $\Sigma^{t,s}$, with signature $(1^t,(-1)^s)$, (which does not have to be  asymptotically flat, or enjoy  any other special property) can be isometrically embedded into a flat N-dimensional pseudo-riemannian space $\mathbb{R}^{t , s}$ with metric 
\be
  \eta =\text{diag}\left(1^t,(-1)^s\right)
 \ee
provided that $n\leq N\leq n(n+1)/2$ as well as $ s\leq S$ and also $ t\leq T$.
\par
It is well known that 4-dimensional constant curvature spaces (de Sitter or anti-de Sitter) can be embedded in a five dimensional ambient flat space $\mathbb{R}^{1,4}$ or else $\mathbb{R}^{2,3}$; that are  their hyperquadrics.
In case $\Sigma^{t,s}$ is Ricci-flat, however, there is the extra condition $N\geq n+2$.
\par
This means that at least six dimensions are neccessary in order to so embed any vacuum solution of Einstein equations, such as Schwarzschild's.
 General Relativity defined out of an embedding has been considered previously in \cite {Regge}; but their aim and techniques were quite different from ours.
 \par
 The structure of this note  is as follows. In section 2 we construct an action out of which we get the energy-momentum tensor and the associated energy for a general spacetime embedded in Minkowski spacetime. In section 3 we apply this idea to Schwarzschild spacetime embedded in $M_6$ and $M_8$. Section 4 is devoted to the study of the embeddings of $dS$ and $AdS$ spacetimes. Finally we conclude with some remarks. 

\section{The action principle for the embedding}

Our aim in this section is to construct an action principle for the embedding of our spacetime, making use of the Gauss-Codazzi equation. 

Let us start by fixing our notation. We consider an embedding of a d-dimensional hypersurface $\Sigma$ into a n-dimensional flat spacetime M with coordinates
\be
x^i\in \Sigma \hookrightarrow z^\m\in M
\ee
We can construct a basis on the tangent space $T(\Sigma)$ as
\be
\t_i^\m\equiv {\pd z^\m\over \pd x^i}
\ee
The induced metric, or first fundamental form, is given by
\be
h_{ij}\equiv \t_i.\t_j
\ee
We have $n-d$ normal vectors defined as
\be
n^A.\t_i=0
\ee
which we assume normalized
\be
n_A^2=(-1)^A
\ee
With our conventions, we have $A=0$ for timelike normals, and $A=1$ for spacelike ones.
The closure relationship means that
\be
h^{ij} \t_i^\m \t_{j\n}+\sum_A (-1)^A n_A^\m n_{A\n}\equiv \left(P_\Sigma\right)^\m_\n+\left(P_{\perp}\right)^\m_\n =\d^\m_\n
\ee
so that the tangent projector is given (for orthonormalized normals) by
\be
(P_\Sigma)^\m_\nu\equiv \d^\m_\n-\sum_A (-1)^A n_A^\m n_{A\n}
\ee
In Appendix \ref{B} we give a different expression valid for unnormalized and unorthonormalized normal vectors.

Then the {\em extrinsic curvature} associated to the embedding is given by
\be
K^A_{ij}\equiv -\t_i^\m \t_j^\n \nabla_\m n^A_\n\equiv  -\t_i^\m \t_j^\n\,{\cal K}_{\m\n}
\ee
where we have defined the ambient tensor
\be
{\cal K}^A_{\m\n}\equiv \nabla_\m n^A_\n
\label{kappa}
\ee
The corresponding trace is given by
\bea
&K=\text{tr}\,\left(P_\Sigma{\cal  K}\right)\equiv P^{\m\n}_\Sigma {\cal K}_{\m\n}\nonumber\\
&K_{ij} K^{ij}=\t_i^\m \t_j^\n {\cal K}_{\m\n} \t^i_\a \t^j_\b {\cal K}^{\a\b}={\cal K}_{\m\n}^2+\left(\text{tr} P_\perp K\right)^2
\eea
and for orthonormalized normals we will have
\be
P_\perp K=0
\ee

Let us consider an action defined in the ambient space, which is essentially equivalent to Einstein-Hilbert. Notice that
the Gauss-Codazzi equation for the embedding $\Sigma \hookrightarrow M_6$ reads
\be
R_{ijkl}^\Sigma=\sum_A (-1)^{\e_{\text{\tiny{A}}}}\left( K^A_{ik}K_{jl}^A-K_{il}^A K^A_{jk}\right)
\ee
where we have used the fact that
\be
R_{\a\b\g\d}=0
\ee
it follows that
\be
R_{jl}^\Sigma=\sum_A (-1)^{\e_{\text{\tiny{A}}}} \sum_i \left( K^A K_{jl}^A-K_{il}^A K^A_{ji}\right)
\ee
\be
R^\Sigma=\sum_A (-1)^{\e_{\text{\tiny{A}}}}\left( K^A K^A-K_{ij}^A K^A_{ji}\right)
\ee
\bea
&S_{\text{\tiny EH}}^\Sigma=-{1\over 2\kappa^2} \int_{\Sigma} d(vol)_\Sigma\,\sum_A (-1)^{\e_{\text{\tiny{A}}}}\left( K^A K^A-K_{ij}^A K^A_{ji}\right)=\nonumber\\&=
-M_p^{d-2} \int_{\Sigma} d(vol)_\Sigma\,\sum_A (-1)^{\e_{\text{\tiny{A}}}}\left( K^A K^A-K_{ij}^A K^A_{ji}\right)
\label{KK}
\eea
Here $\kappa$ has mass dimension $\frac{2-d}{2}$. 

\par

The idea of this paper is to consider the extension of the Einstein-Hilbert action to the ambient Minkowski spacetime, where we have a well defined notion of energy, and see if at the end we can somehow project the results back into the hypersurface. This is just the action \eqref{KK} before projecting into the submanifold, that is, using the ambient tensor $\cal{K}_{\mu \nu}$ \eqref{kappa} 

\be
S \equiv M_p^{n-2} \int d^n z  \ \sum_{A} \bigg\{\left(\pd_\m n_A^\m\right)^2-\left(\pd_\m n^A_\l\right)^2\bigg\}
\ee

We have arranged the normal vectors to be dimensionless, and we have defined $M_p$ as the Planck mass. In this way, the normal vectors $n_\m(z)$ are considered to be defined in the whole n-dimensional Minkowski space. 
The EM read\footnote{From now on, we will omit the index $A$ denoting the different normal vectors. }
\be
\Box n_\l-\pd_\l \left(\pd_\a n^\a\right)=0
\ee
conveying the conservation of the momentum 
\be
\pi_{\a\b}\equiv M_p^{n-2}\left(\pd_\a n_\b-\eta_{\a\b}\pd_\l n^\l\right)
\ee
The canonical energy-momentum tensor is given by
\be
T_{\m\n}=2M_p^{n-2}\left(\eta^\r_{~\m} \pd_\a n^\a-\pd_\m n^\r\right)\pd_\n n_\r-L\eta_{\m\n}
\label{EMtensor}
\ee
in such a way that
\bea
 T_{00}&=M_p^{n-2}\left[2\pd_\a n^\a \pd_0 n_0-2\pd_0 n^\a\pd_0 n_\a-\left(\pd_\a n^\a\right)^2+\left(\pd_\m n_\l\right)^2\right]=\nonumber\\
 &=- M_p^{n-2}\left(\dot{n}^2-(\nabla n_0)^2 +\partial_i n^i \partial_j n^j - \partial_i n_j\partial^i n^j \right)
 \eea

The momenta are given by
\bea
&\pi_0=2M_p^{n-2}\nabla n\nonumber\\
&\pi=- 2 M_p^{n-2}\dot{n}
\eea
and the Hamiltonian
\bea
& H= M_p^{n-2}\left(-{\pi_0^2\over 4}- {\pi^2\over 4} +\left(\nabla n_0\right)^2 + \partial_j n_i \partial^j n^i  \right)
\eea
exactly coincides with the {\em canonical} $T_{00}$.

We are also going to be interested in general in the energy as measured by an observed lying in an hypersurface
\be
f(z)=0
\ee
this is given by
\be
P^\m_f\equiv T^{\m\n} N_\n
\ee
where
\be
N_\m\equiv {\pd_\m f\over \sqrt{\eta^{\a\b}\pd_\a f\pd_\b f}}
\ee



\section{ Schwarzschild as a brane}
Here we use the ideas of the previous section applied to 4-dimensional Schwarzschild spacetimes.
We shall dub, somewhat whimsically, {\em brane} to the Schwarzschild spacetime itself, and {\em ambient spacetime} to the target Minkowski space in which the brane is embedded.
\subsection{Fronsdal's embedding in $M_6$}
The most famous embedding of Schwarzschild in flat spacetime is Fronsdal's \cite{Fronsdal} where it is considered as a brane $\Sigma$ embedded  in $z\in M_6$ as
	 \bea
	& z_0=2 r_s\sqrt{1-{r_s\over r}}\,\sinh\,{t\over 2 r_s}\nonumber\\
	 & z_1=2 r_s\sqrt{1-{r_s\over r}}\,\cosh\,{t\over 2 r_s}\nonumber\\
 &z_2 =f(r)\equiv  \int dr \sqrt{\dfrac{(r_s r^2 + r_s^2 r + r_s^3)}{r^3}}\Longrightarrow r=r(z_2) \nonumber \\
 &z_3^2+z_4^2+z_5^2= r^2
\eea
Kruskal's maximal analytic extension is obtained by considering the two branches of solutions of the above, as was already mentioned by Fronsdal in his original paper. 

Our philosophy here is to consider the whole Schwarzschild hypersurface as a brane embedded in $M_6$,
$\Sigma\hookrightarrow M_6$. The brane can be characterized as
\bea
&z_1^2-z_0^2= 4 r^2_s\left(1-{r_s\over r(z_2)}\right)\nonumber\\
&z_3^2+z_4^2+z_5^2=r^2(z_2)
\eea
This particular embedding is not however optimal for our purposes, because of the difficulties coming from the integral defining $z_2$ and the explicit definition of $r=r(z_2)$.
\subsection{Embedding in $M_{8}$}
Let us generalize Frondal's embedding to the 8-dimensional Mikowski spacetime, $M_8$,
\bea
Z_0&=2r_s\sqrt{1-\frac{r_s}{r}}\sinh \frac{t}{2r_s}\nonumber\\
Z_1&=2r_s\sqrt{1-\frac{r_s}{r}}\cosh \frac{t}{2r_s}\nonumber\\
Z_2&=f(r)=2\sqrt{r_sr}\nonumber \\
Z_3&=g(r)=r_s\log \frac{r}{r_s}\nonumber \\
Z_4&=h(r)=-2r_s\sqrt{\frac{r_s}{r}}\nonumber \\
Z_5&=r\sin \theta \sin \phi \ ,  \hspace{.5cm} Z_6=r\sin \theta \cos \phi  \ ,  \hspace{.5cm} Z_7=r\cos \theta\eea
The advantage of this embedding is the freedom of choosing the functions $Z_2$,$Z_3$ and $Z_4$, which will yield the spatial part of the 4-dimensional Schwarzschild metric, and therefore the possibility of analitically obtaining $r = r(Z_i)$. Note that when $r_s\rightarrow 0$, the trivial embedding of $M_4$ in $M_8$ is recovered.

Equivalently this can be seen as brane embedded in $M_8$ defined by
\bea
&Z_1^2-Z_0^2- 4 r_s^2\left(1-{r_s\over r}\right) = 0\nonumber\\
&Z_2-2\sqrt{r_sr}=0\nonumber\\
&Z_3 -r_s\log \frac{r}{r_s}=0\nonumber\\
&Z_4+2r_s\sqrt{\frac{r_s}{r}} = 0
\label{embeddingM8}
\eea
where we take $r=\sqrt{Z_5^2 + Z_6^2 + Z_7^2}$. 
With this description, it is straightforward to obtain the (unothonormalized) normal vectors to the hypersurface 

\bea
{(n_1)}_\mu &= \frac{1}{2r_s}\left(2 Z_0,-2 Z_1,0,0,0,\frac{4r_s^3}{r^2}r_5,\frac{4r_s^3}{r^2}r_6,\frac{4r_s^3}{r^2}r_7\right)\nonumber\\
{(n_2)}_\mu &= \left(0,0,1,0,0,-\sqrt{\frac{r_s}{r}}r_5,-\sqrt{\frac{r_s}{r}} r_6,-\sqrt{\frac{r_s}{r}}r_7 \right)\nonumber\\
{(n_3)}_\mu &= \left(0,0,0,1,0,-\frac{r_s}{r}r_5,-\frac{r_s}{r} r_6,-\frac{r_s}{r} r_7 \right)\nonumber\\
{(n_4)}_\mu &= \left(0,0,0,0,1,-\sqrt{\frac{r^3_s}{r^3}}r_5,-\sqrt{\frac{r^3_s}{r^3}}r_6,-\sqrt{\frac{r^3_s}{r^3}}r_7\right)
\label{normals}
\eea
The $r_i$ notation stands for $r_i=\frac{\partial r}{\partial Z_i}=\frac{Z_i}{\sqrt{Z_5^2 + Z_6^2 + Z_7^2}}$.

These vectors satisfy the equations of motion
\be
\Box n^\m - \partial^\m \partial_\nu n^\nu = 0
\ee
Therefore, with these normal vectors, we can obtain the explicit expression for the canonical energy momentum tensor,which will be automatically conserved in the usual sense (its explicit form is shown  in Appendix \ref{A}).  

A natural way to restrict this tensor to the Schwarzschild brane is to compute the pullback 
\begin{equation}\label{key}
T^\Sigma_{ij}=\tau_i^\mu \tau_j^\nu T_{\mu\nu}
\end{equation}
with the tangent vectors defined as
\begin{equation}
\tau_i^\mu=\frac{\partial Z^\mu}{\partial x^i}
\end{equation}
where $x^i$ are the coordinates on Schwarzschild, $(t,r,\theta,\phi)$. In this way, we obtain a quite large tensor, whose $00$ component is nevertheless very simple 
\begin{equation}
T^\Sigma_{00}= M_p^6 \left( \frac{2 r_s^2}{r^4}+\frac{2 r_s^3}{r^5}+\frac{20 r_s^4}{r^6}-\frac{24 r_s^5}{r^7} \right)
\end{equation}
From these expressions one can already foresee that there is no way that a linear term in $r_s$ can appear in the energy (the ADM mass is of course proportional to $M_p^2 r_s$). On hindsight this is just a consequence of the action used as the starting point, which is {\em quadratic} in the extrinsic curvature; whereas the boundary action used in deriving the ADM mass is linear in the extrinsic curvature. The extrinsic curvature in this case takes the form
\be
{\cal{K}} = \dfrac{2}{r_s} + 3 \dfrac{\sqrt{r r_s}}{r^2} + \dfrac{r_s}{r^2} + \dfrac{r_s}{2r^2}\sqrt{\dfrac{r_s}{r}}
\ee
Due to cancelations the leading term in the energy momentum tensor is quadratic in $r_s$.
\par
The next step will be to integrate the energy density over the brane, so that we get the total energy of the 4-dimensional Schwarzschild spacetime. We elaborate on this idea in Appendix \ref{C}. 

\section{Conclusions}
The initial purpose of our work was to try to find a more intrinsic notion of quasi-local energy, in order to get away from the need of withdrawing the zero-point energy that appears in the usual definitions of QLE, either via an isometric embedding in $\mathbb{R}^3$ \cite{Brown:1992br} or else in $M_4$ \cite{Wang}. The idea was to compute the QLE of the full spacetime, defining it as a brane embedded in higher dimensional Minkowski spacetime.

\par
In practice, we have found it difficult even to reproduce the well known result of the Schwarzschild ADM mass. We have written this note to inform the curious reader that some natural ideas do not seem to work properly. As the dictum says, 
{\em the road to hell is paved with good intentions.}
\section{Acknowledgements}

 GMB is supported by the project FPA2015-65480-P and RSG is supported by the Spanish FPU Grant No. FPU16/01595. This work has received funding from the European Unions Horizon 2020 research and innovation programme under the Marie Sklodowska-Curie grants agreement No 674896 and No 690575. We also have been partially supported by FPA2016-78645-P(Spain), COST actions MP1405 (Quantum Structure of Spacetime) and  COST MP1210 (The string theory Universe). This work is supported by the Spanish Research Agency (Agencia Estatal de Investigacion) through the grant IFT Centro de Excelencia Severo Ochoa SEV-2016-0597.
\appendix

\section{Energy momentum tensor of $Sch_4 \hookrightarrow M_8$}
\label{A}
The  explicit expression for the only non zero components read

	\bea
	&\hat{T}_{00}= -\hat{T}_{11} =	\frac{2 r_s^2 \left(r^2+2 r_s r+12 r_s^2\right)}{r^6}  \nonumber \\
	& \hat{T}_{22}= \hat{T}_{33} =	T_{44} = \frac{2 \left(r^6-r_s^4 r^2-2 r_s^5 r-12 r_s^6\right)}{r^6 r_s^2}  \nonumber \\
	& \hat{T}_{55}= \frac{2 r^9-7 r^4 r_s^3 \left(2 Z_5^2-Z_6^2-Z_7^2\right)+8 r r_s^6 \left(Z_5^2-2 \left(Z_6^2+Z_7^2\right)\right)+2 r r_s^4 \left(Z_5^4-\left(Z_6^2+Z_7^2\right)^2\right)}{r^9 r_s^2} \nonumber \\
	&+\dfrac{r_s^5 \left(2 Z_5^4-\left(Z_6^2+Z_7^2\right) Z_5^2-3 \left(Z_6^2+Z_7^2\right)^2\right)}{r^9 r_s^2}  \nonumber \\
	& \hat{T}_{66} = \frac{2 r^9+7 r_s^3 \left(Z_5^2-2 Z_6^2+Z_7^2\right) r^4-2 r_s^4 \left(Z_5^4+2 Z_7^2 Z_5^2-Z_6^4+Z_7^4+r_s^2 \left(8 Z_5^2-4 Z_6^2+8 Z_7^2\right)\right) r}{r^9 r_s^2} \nonumber \\
	&-\dfrac{r_s^5 \left(3 Z_5^4+\left(Z_6^2+6 Z_7^2\right) Z_5^2-2 Z_6^4+3 Z_7^4+Z_6^2 Z_7^2\right)}{r^9  r_s^2} \nonumber \\
	& \hat{T}_{77}= \frac{2  r^9+7 r_s^3 \left(Z_5^2+Z_6^2-2 Z_7^2\right) r^4-2 r_s^4 \left(Z_5^4+2 Z_6^2 Z_5^2+Z_6^4-Z_7^4+r_s^2 \left(8 Z_5^2+8 Z_6^2-4 Z_7^2\right)\right) r}{r^9 r_s^2} \nonumber \\
	&-\dfrac{r_s^5 \left(3 Z_5^4+\left(6 Z_6^2+Z_7^2\right) Z_5^2+3 Z_6^4-2 Z_7^4+Z_6^2 Z_7^2\right)}{r^9 r_s^2}  \nonumber \\
	& \hat{T}_{56} =  \frac{r_s \left(4 r_s r^3-21 r^4+24 r r_s^3+5 r^2 r_s^2\right) Z_5 Z_6}{r^9} \nonumber \\
	& \hat{T}_{57} =  \frac{r_s \left(4 r_s r^3-21 r^4+24 r r_s^3+5 r^2 r_s^2\right) Z_5 Z_7}{\left(r^2\right)^{9/2}} \nonumber \\
	& \hat{T}_{67} = \frac{r_s \left(4 r_sr^3-21 r^4+24 r r_s^3+5 r^2 r_s^2\right) Z_6 Z_7}{r^9}
	\eea

\section{Projectors}
\label{B}

We  claim that we can afford to use unnormalized normals provided we redefine the tangent projector.
For unitary, but non-orthogonal normals ($n_0^2=1\,,n_1^2=-1\,,n_0.n_1\neq 0$). The tangent projector reads
\be
P^\a_\b=\d^a_\b -{1\over 1+(n_1.n_0)^2}\bigg\{n_0^\a n^0_\b-n_1^\a n^1_\b+(n_0.n_1)\left( n_0^\a n^1_\b+n_1^\a n^0_\b\right)\bigg\}
\ee
This projector generalizes trivially to the case of more than two normal vectors.

On the other hand, the tangent projector with unnormalized normals (one spacelike and the other timelike)  
\be
n_0^2\equiv |n_0|^2
\ee
and
\be
n_i^2=-|n_i|^2
\ee 
can also be computed as 
\be
P^\a_\b=\d^\a_\b-{|n_0|^2|n_1|^2\over |n_0|^2|n_1|^2+(n_0.n_1)^2}\bigg\{{n_0^\a n^0_\b\over |n_0|^2}-{n_1^\a n^1_\b\over |n_1|^2}+{n_0.n_1\over |n_0|^2|n_1|^2}\left(n_0^\a n^1_\b+n_1^\a n^0_\b\right)\bigg\}
\ee
This tangent projector can be used to project any quantity computed using unorthonormalized normal vectors, so that we could in principle restrict our computations to the Schwarzschild brane.


%
%
\section{Integration}
\label{C}
\
Once we have obtained the EM tensor of Appendix \ref{A}, there are different definitions of energy that we can construct. For example, we can focus on the energy that will be measured by an observer at constant Minkowski time $Z_0$, with unit normal vector $u^\m = (1, \vec{0})$. Then, the value of the $00$ component of the EM tensor yields
\be
T_{00}^u \equiv M_p^6\hat{T}^u_{00}=M_p^{6}\left(\frac{2 r_s^2}{r^4}+\frac{4  r_s^3}{r^5}+\frac{24 r_s^4}{r^6}\right)
\label{00}
\ee

Another possibility is to define the energy as the one measured by an observer at constant Schwarzschild time $t$.  This corresponds in the ambient space to
\be
\dfrac{Z_0}{Z_1}= C \ \rightarrow \ v_\mu = \dfrac{1}{\sqrt{1-C^2}}(1,-C,0,0,0,0)
\ee
where $C = \tanh (t/2r_s)$ is constant. This yields again the same energy density
\be
v^\m v^\n T_{\m \n} =M_p^{6}\left(\frac{2 r_s^2}{r^4}+\frac{4  r_s^3}{r^5}+\frac{24 r_s^4}{r^6}\right)
\label{00v}
\ee

Now we want to integrate  the first expression at constant $Z_0$, and the second one at constant $t$. We could integrate in the full ambient space, but the integrals will be divergent. It makes sense to restrict the integration to the brane introducing the delta functions

\bea
E&=M_p\int dZ_5  dZ_6  dZ_7  dZ_0 dZ_1  dZ_2  dZ_3  dZ_4 \hat{ T}_{00}^u[Z_5,Z_6,Z_7]  \delta (Z_2 - f(r)) \ \delta(Z_3 - g(r)) \  \delta(Z_4 - h(r))  \nonumber \\
&\delta \left(Z_1^2 - Z_0^2 - 4 r_s^2\left(1 - \dfrac{r_s}{\sqrt{Z_5^2 + Z_6^2 + Z_7^2}}\right) \right) \delta (Z_0 - C)
\eea
The integration in the $Z_2$, $Z_3$ and $Z_4$ coordinates is trivial. Using the properties of the Dirac delta, we can write the delta function involving $Z_0$ and $Z_1$ as
\bea
E&=M_p\int dZ_5  dZ_6  dZ_7  dZ_1   \ \hat{T}_{00}[Z_5,Z_6,Z_7] \ 
 \dfrac{1}{2 \sqrt{ C^2 + 4 r_s^2 \left(1 - \dfrac{r_s}{\sqrt{Z_5^2 + Z_6^2 + Z_7^2}}\right)}} \nonumber \\
 &\left\lbrace \delta \left(Z_1 + \sqrt{ C^2 + 4 r_s^2 \left(1 - \dfrac{r_s}{\sqrt{Z_5^2 + Z_6^2 + Z_7^2}}\right)} \right)+ \delta \left(Z_1 - \sqrt{ C^2 + 4 r_s^2 \left(1 - \dfrac{r_s}{\sqrt{Z_5^2 + Z_6^2 + Z_7^2}}\right) }\right) \right\rbrace 
\eea
therefore
\be
E =M_p \int dZ_5  dZ_6  dZ_7    \hat{ T}_{00}[Z_5,Z_6,Z_7] \ 
\dfrac{1}{ \sqrt{ C^2 + 4 r_s^2 \left(1 - \dfrac{r_s}{\sqrt{Z_5^2 + Z_6^2 + Z_7^2}}\right)}} 
\ee

Using spherical coordinates for $Z_5,Z_6,Z_7$

\bea
E &= M_p\int dr \ d\theta \ d\phi \ r^2 \sin \theta   \ \left(\frac{2 r_s^2}{r^4}+\frac{4  r_s^3}{r^5}+ \frac{24 r_s^4}{r^6} \right) \ 
\dfrac{M_p}{ \sqrt{ C^2 + 4 r_s^2 \left(1 - \dfrac{r_s}{r}\right)}} \nonumber \\
&=4 \pi  M_p \  \sqrt{-\frac{4 r_s^3}{r}+4 r_s^2+C^2} \left( \frac{  131 r_s^4+53 r_s^2 C^2+6 C^4}{15 r_s^5} + \dfrac{58 r_s^5+ 12 r_s^3 C^2}{15 r r_s^2}+ \dfrac{36 r_s }{15 r^2 }\right)
\eea
For $r \rightarrow \infty$ we get
\be
E \simeq M_p\left(-\frac{8 \pi  r_s^2}{r \ \sqrt{4 r_s^2+C^2}}+\frac{4 \pi \ \sqrt{4 r_s^2+C^2} \left(131 r_s^4+53 r_s^2 C^2+6 C^4\right) }{15 r_s^5} \right)+ O\left(\frac{1}{r}\right)^2
\ee
\par

If we do the same for the observer at rest in Schwarzschild's time we have 

\bea
E^v&= M_p\int dZ_5  dZ_6  dZ_7  dZ_0 dZ_1  dZ_2  dZ_3  dZ_4 \hat{ T}_{vv}[Z_5,Z_6,Z_7]  \delta (Z_2 - f(r)) \ \delta(Z_3 - g(r)) \  \delta(Z_4 - h(r))  \nonumber \\
&\delta \left(Z_1^2 - Z_0^2 - 4 r_s^2\left(1 - \dfrac{r_s}{\sqrt{Z_5^2 + Z_6^2 + Z_7^2}}\right) \right) \delta (Z_0 - C Z_1)
\eea
so that integrating in $Z_0,Z_2,Z_3,Z_4$
\bea
E^v&= M_p\int dZ_5  dZ_6  dZ_7  dZ_1   \hat{ T}_{vv}[Z_5,Z_6,Z_7] 
&\delta \left((1-C^2)Z_1^2 - 4 r_s^2\left(1 - \dfrac{r_s}{\sqrt{Z_5^2 + Z_6^2 + Z_7^2}}\right) \right) 
\eea
After integrating in $Z_1$ we get
\bea
E^v&= M_p\int dZ_5  dZ_6  dZ_7    \hat{ T}_{vv}[Z_5,Z_6,Z_7]  \dfrac{1}{2 r_s (1-C^2)^{1/2} \sqrt{ \left(1 - \dfrac{r_s}{\sqrt{Z_5^2 + Z_6^2 + Z_7^2}}\right)}} \nonumber \\
\eea
Finally, we use spherical coordinates, and the energy measured by this observer reads
\be
E^v = 8 \pi M_p \  \sqrt{1-\frac{r_s}{r}} \left( \frac{131 r^2+58 r r_s+36 r_s^2}{15 r^2 \ \sqrt{1-C^2}} \right)
\ee
and for $r \rightarrow \infty$
\be
E^v \simeq \dfrac{M_p}{\sqrt{1-C^2}} \left( \frac{1048 \pi }{15  }-\frac{4 \pi  r_s}{r } \right) +O\left(\frac{1}{r}\right)^2
\ee

It seems that even for an observer at rest in Schwarzschild, we cannot avoid a dependence in the constant time upon integration. This is due to the fact that constant Schwarzschild time mixes the $Z_0$ and $Z_1$ coordinates of the ambient space, as it does one of the defining equations of the hypersurface. 
\par
From this point of view one could think that it would be more natural to work in synchronous coordinates, which have been worked out for the Schwarzschild metric by Lema\^itre \cite{Lemaitre}
\be
ds^2= d\tau^2-{r_s\over r}d\r^2-r^2d\Omega_2^2
\ee
where
\be
r\equiv \left({3\over 2}\left(\r-\t\right)\right)^{2/ 3} r_s^{1/ 3}
\ee
This will yield $Z_0 = \tau$ but there will be a $\tau$ dependence in the other coordinates so that the problem with the integration at constant time will still be there. 




\newpage

	\end{document}